\documentclass[conference]{IEEEtran}
\ifCLASSINFOpdf
  \usepackage[pdftex]{graphicx}
\else
 \usepackage[dvipdfmx]{graphicx}
\fi

\usepackage{url}
\usepackage{tabularx}
\usepackage{booktabs}
\usepackage{textcomp}

\begin{document}
\title{Software Defined Media:\\Virtualization of Audio-Visual Services}


%\author{\IEEEauthorblockN{Manabu Tsukada}
%\IEEEauthorblockA{School of Electrical and\\Computer Engineering\\
%Georgia Institute of Technology\\
%Atlanta, Georgia 30332--0250\\
%Email: http://www.michaelshell.org/contact.html}
%\and
%\IEEEauthorblockN{Homer Simpson}
%\IEEEauthorblockA{Twentieth Century Fox\\
%\and
%\IEEEauthorblockN{James Kirk\\ and Montgomery Scott}
%\IEEEauthorblockA{Starfleet Academy\\
%San Francisco, California 96678--2391\\



\author{\IEEEauthorblockN{Manabu Tsukada\IEEEauthorrefmark{1},
  Keiko, Ogawa\IEEEauthorrefmark{2},
  Masahiro Ikeda\IEEEauthorrefmark{3},
  Takuro Sone\IEEEauthorrefmark{3},
  Kenta Niwa\IEEEauthorrefmark{4},  \\
  Shoichiro Saito\IEEEauthorrefmark{4},
  Takashi Kasuya\IEEEauthorrefmark{5},
  Hideki Sunahara\IEEEauthorrefmark{2}, and
  Hiroshi Esaki\IEEEauthorrefmark{1}
}
\IEEEauthorblockA{\IEEEauthorrefmark{1}
Graduate School of Information Science and Technology, The University of Tokyo\\
Email: \textit{tsukada@hongo.wide.ad.jp, hiroshi@wide.ad.jp}}
\IEEEauthorblockA{\IEEEauthorrefmark{2}Graduate School of Media Design, Keio University /
%Email: \textit{\{keikoogawa, suna\}@kmd.keio.ac.jp}}
Email: \textit{keikoogawa@kmd.keio.ac.jp, suna@wide.ad.jp}}
\IEEEauthorblockA{\IEEEauthorrefmark{3}Yamaha Corporation /
Email: \textit{\{masahiro.ikeda, takurou.sone\}@music.yamaha.com}}
\IEEEauthorblockA{\IEEEauthorrefmark{4}NTT Media Intelligence Laboratories /
Email: \textit{\{niwa.kenta, saito.shoichiro\}@lab.ntt.co.jp}}
\IEEEauthorblockA{\IEEEauthorrefmark{5}Takenaka Corporation /
Email: \textit{kasuya.takashi@takenaka.co.jp}}
}

\maketitle

\begin{abstract}

Internet-native audio-visual services are witnessing rapid development. Among these services, object-based audio-visual services are gaining importance. In 2014, we established the Software Defined Media (SDM) consortium to target new research areas and markets involving object-based digital media and Internet-by-design audio-visual environments.
In this paper, we introduce the SDM architecture that virtualizes networked audio-visual services along with the development of smart buildings and smart cities using Internet of Things (IoT) devices and smart building facilities.
Moreover, we design the SDM architecture as a layered architecture to promote the development of innovative applications on the basis of rapid advancements in software-defined networking (SDN).
Then, we implement a prototype system based on the architecture, present the system at an exhibition, and provide it as an SDM API to application developers at hackathons.
Various types of applications are developed using the API at these events.
An evaluation of SDM API access shows that the prototype SDM platform effectively provides 3D audio reproducibility and interactiveness for SDM applications.

\end{abstract}

% Media Networking; Media application; Audio Visual; Smart building; 3D Media

\IEEEpeerreviewmaketitle

\section{Introduction}

With the proliferation of the Internet, users are increasingly able to enjoy media generation services, such as Vocaloid\texttrademark, and media sharing services such as YouTube, Facebook, and Nico Nico Douga.
In these services, data is generated, processed, transmitted, and shared in a digitally native manner on the Internet without analog-to-digital conversion.
Recently, the object-based approach has gained importance, whereby media data is decoupled from audio-visual data and 3D meta-data, which represents multiple objects that exist in 3D space, and transmitted to remote locations for the reproduction of 3D space.
This allows flexible and possibly interactive reproduction that adapts to the receiver's configuration, including head-mounted display (HMD), 3D TV, and 3D audio systems.

Object-based audio systems for reproducing 3D sound are commercially used in cinemas, home theatres, and broadcasting TV systems (e.g., Dolby Atmos\cite{Dolby2015} and DTS:X~\cite{DTSX}).
Furthermore, researchers are developing free-viewpoint video systems that employ object-based systems~\cite{Smolic2006, Sankoh2012}, whereby the 3D models of visual objects are extracted from multiple field-installed cameras.

We believe that innovative applications will emerge from the fusion of object-based audio and video systems, including new interactive education systems and public viewing systems. In 2014, we established the Software Defined Media (SDM) consortium\footnote{\url{http://sdm.wide.ad.jp/}} to target new research areas and markets involving object-based digital media and Internet-by-design audio-visual environments.
We design SDM along with the development of smart buildings and smart cities using Internet of Things (IoT) devices and smart building facilities.
Moreover, we define audio-visual devices as smart building facilities for better integration into the smart buildings.
SDM virtualizes the networked audio-visual infrastructure and enables SDM applications to flexibly access audio-visual services on the basis of rapid advancements in software-defined networking (SDN)~\cite{Feamster2013}.

%- Smart city / Smart building \\
% - Audio-visual infrastructure in building facility \\
%- Virtualization for flexible audio-visual service \\
%- SDN / NFV \\

This paper discusses the activities and achievements of the SDM consortium.
Section~\ref{sec:related} reviews the related studies.
Section~\ref{sec:goals} describes the motivations and goals of SDM.
Section~\ref{sec:arch} introduces the SDM architecture that realizes flexible audio-visual services based on object-based media.
Section~\ref{sec:prototype} presents an overview of the SDM prototype system.
Section~\ref{sec:applications} summarizes the SDM platform presented at exhibitions and provided at hackathons.
Section~\ref{sec:evaluation} describes the evaluation of the SDM prototype systems.
Finally, Section~\ref{sec:conclusion} concludes the paper and explores directions for future work.

\section{Related Work} \label{sec:related}

Sound reproduction with spatial expression began with the development of stereo sound (two-channel) and continued with the development of surround sound (multi-channel).
22.2 Multichannel Sound~\cite{hamasaki201122} introduced height expression for 3D sound reproduction.
Super Hi-Vision \cite{Nakasu2012} employs 22.2 multichannel surround sound with the world’s first High-Efficiency Video Coding (HEVC) encoder for 8K Ultra HDTV.
Channel-based systems reproduce sound from the information assigned to each channel, whereas object-based audio systems render sounds based on software processing while considering the 3D locations of the sound objects as well as the positions of the speakers.
Besides Dolby Atmos\cite{Dolby2015} and DTS:X~\cite{DTSX}, which have been deployed in theatres, an object-based audio format is being standardized by the ISO/IEC Moving Picture Experts Group (MPEG) as MPEG-H for consumer audio including broadcasting~\cite{Herre2015}.
Flexibly focusing on a sound source using signal processing with a microphone array has been investigated as an audio recording technique~\cite{Johnson1992, Niwa2013}.
Moreover, researchers have investigated 3D model extraction from multiple field-installed cameras as a video recording technique~\cite{Smolic2006, Sankoh2012}.

The audio-visual infrastructure installed in a building, including speakers, microphones, displays, and cameras, is considered as a building facility interconnected by an IP network. Recent advancements in smart buildings include facilities such as heating, ventilation, and air conditioning (HVAC) as well as lighting systems managed by intelligent controllers connected to an IP network using standard protocols such as IEEE 1888~\cite{IEEE1888-2016}.

\section{Motivations and Goals}\label{sec:goals}

The notion of Software Defined Media (SDM) has arisen from IP-based media networks. SDM is an architectural approach to media as a service by virtualization and abstraction of networked media infrastructure.
SDM allows application developers to manage audio-visual services through the abstraction of lower-level functionalities. This is achieved by decoupling the software that makes decisions about how to render the audio-visual services from the underlying infrastructure that inputs and outputs the sound. Such decoupling enables a system to configure flexible and intelligent audio-visual services to match an entertainment program with the rendering.

The goals of SDM are as follows:

\begin{itemize}
  \item \textbf{Software-programmable environment of 3D audio-visual services:}
At present, audio-visual systems are mainly hardware-based dedicated systems that fulfill different performance requirements for applications in theatres, museums, classrooms, and so on.
  SDM aims to change such dedicated systems into flexible systems via software rendering. The software manages the audio-visual objects in the virtual 3D space and renders images and sounds suitable for the rendering environments.
  By reconfiguring the software, many applications can be realized with a general-purpose audio-visual infrastructure in the rendering environment.

  \item \textbf{Mixing of 3D audio-visual objects from multiple sources:}
  SDM aims to achieve receiver-side mixing of audio-visual objects from multiple sources, including the Internet and broadcast systems, for both real-time and archived content.
SDM realizes flexible rendering of images and sounds that adapt to the requirements of the audience.

  \item \textbf{Augmented audio-visual effect via software rendering:}
  SDM exchanges multiple audio-visual objects between remote locations.
  Some exchanged objects are recorded from the real space, while other objects may be virtually created to add the desired effects.
  SDM aims to create augmented audio-visual effects by mixing real and virtual audio-visual objects to provide an enriched experience to the audience.

  \item \textbf{Focusing on audience interests:}
  SDM provides means for users to interactively provide feedback regarding their interests to the software systems for reproduction management, content source selection, and so on.
  Moreover, SDM connects spatially/temporally distributed audiences having similar interests and enables them to interact.

\end{itemize}

% ------------------
\section{SDM Architecture}\label{sec:arch}
\begin{figure}[ht]
  \centering
  \includegraphics[width=\columnwidth]{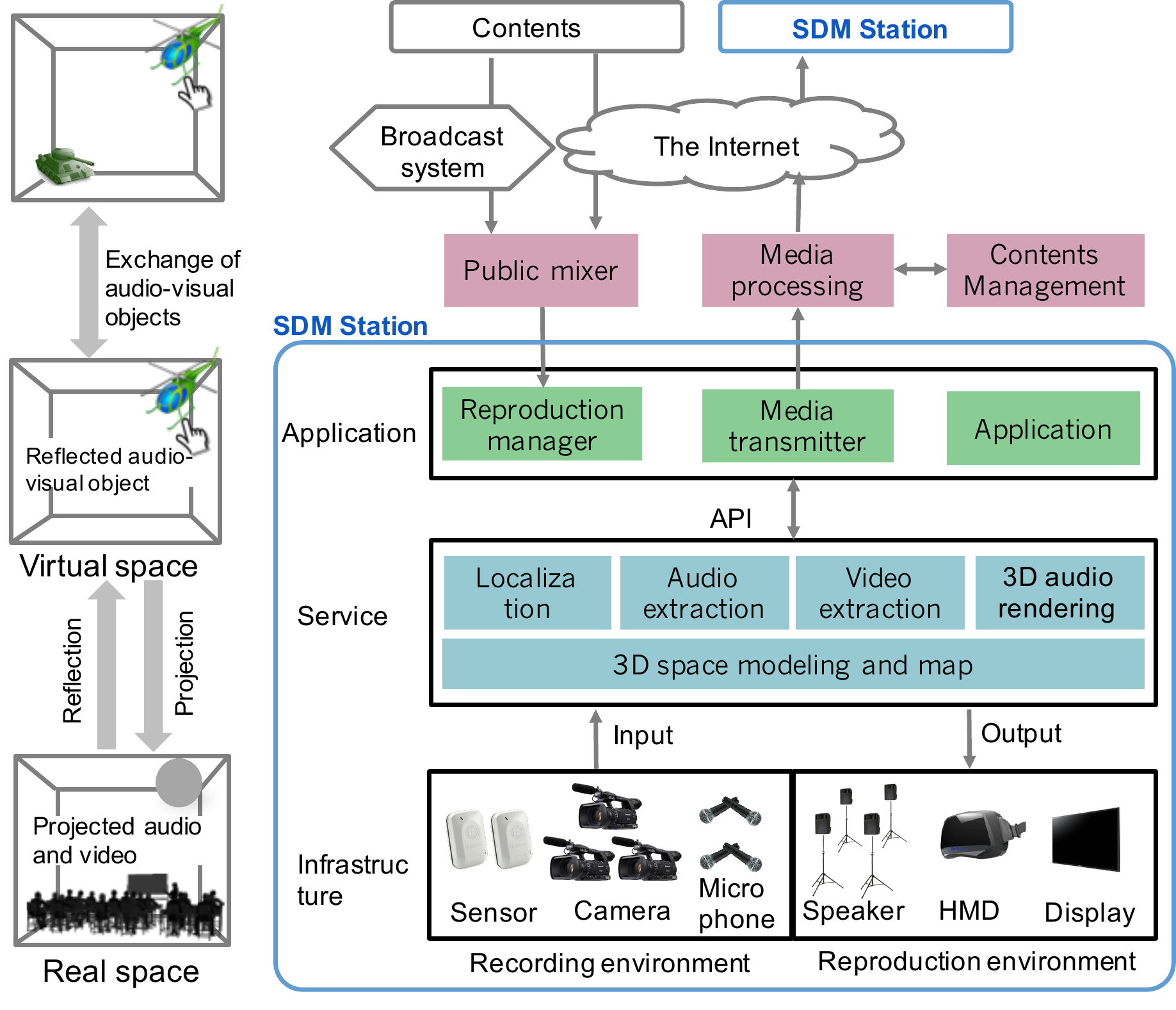}
  \caption{SDM architecture}
  \label{fig:sdm-architecture}
\end{figure}

Fig.~\ref{fig:sdm-architecture} shows the SDM architecture that realizes the goals described in the previous section.

As shown on the left-hand side of Fig.~\ref{fig:sdm-architecture}, SDM interprets the objects in the \textit{\textbf{real space}} and \textit{\textbf{reflects}} them three-dimensionally as audio-visual objects in the \textit{\textbf{virtual space}}. The audio-visual objects are exchanged between remote locations and processed in various applications. The results are \textit{\textbf{projected}} from the virtual space to the real space for the rendering of sounds and images.
The \textit{\textbf{SDM station}} is the most basic unit of SDM for linking the virtual space and the real space.
An SDM station is divided into three layers, as shown on the right-hand side of Fig.~\ref{fig:sdm-architecture}.

The \textit{\textbf{infrastructure layer}} includes input devices that record and interpret the targets three-dimensionally in the real space, e.g., sensors, cameras, and microphones, as well as output devices that render the images and sounds from the audio-visual objects, e.g., speakers, HMDs, and displays.
The architecture is designed to be extensible in anticipation of future innovative devices.
Moreover, we design the architecture to accommodate cases in which the recording environment overlaps with the rendering environment in the real space.

The \textit{\textbf{service layer}} abstracts the functionalities of the infrastructure layer and provides SDM services to applications via an Application Programming Interface (API). This layer defines the 3D model of the space as the fundamental element for service management and provides various services, such as localization, audio extraction, video extraction, and 3D audio rendering.
The API is extensible and flexible, enabling developers to realize creative applications.
\begin{figure*}[hbt]
  \begin{minipage}{0.69\linewidth}
  \includegraphics[width=1\linewidth]{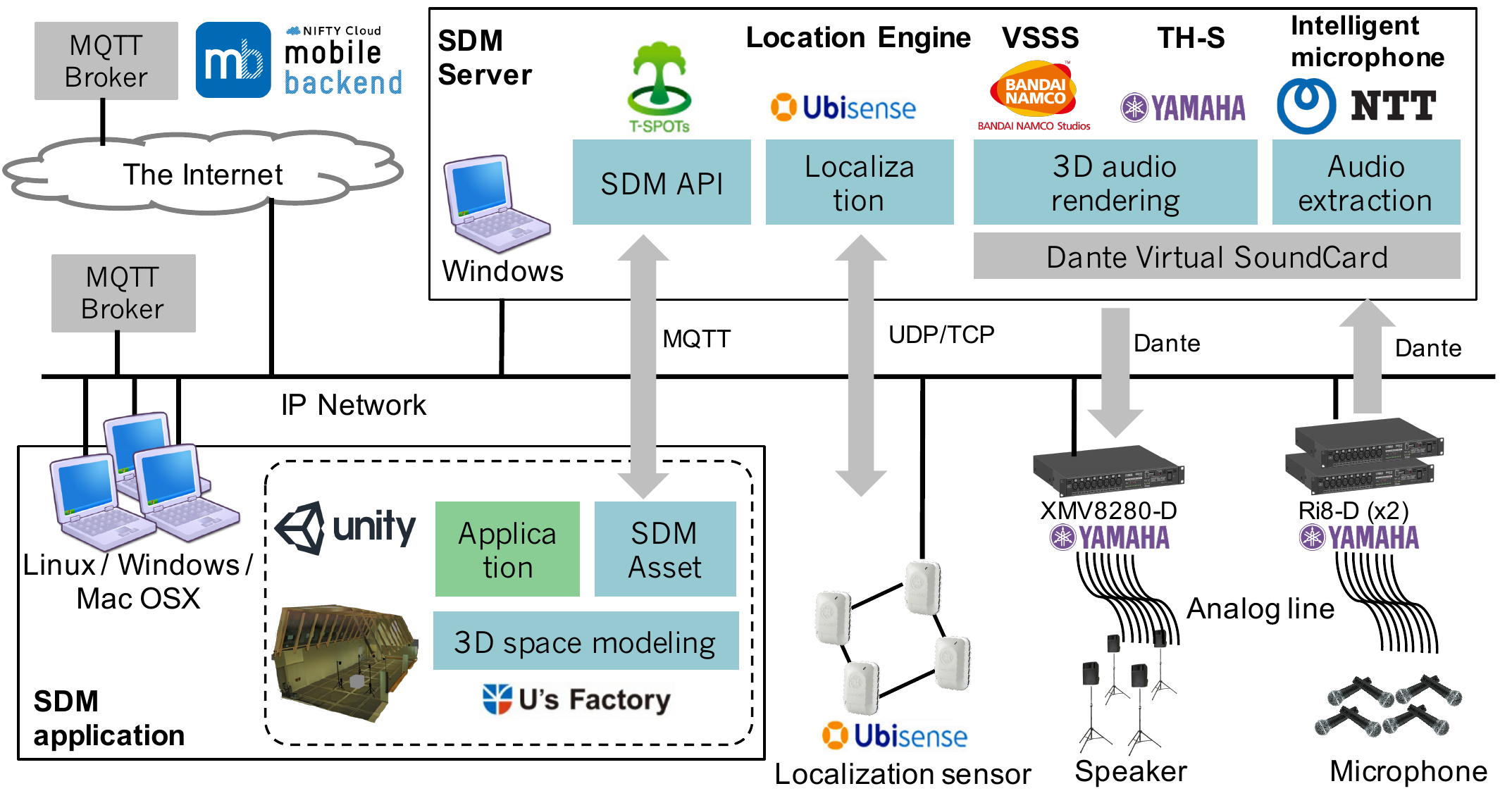}
  \caption{Overview of SDM prototype system}
  \label{fig:prototype-system}
  \end{minipage}
  \begin{minipage}{0.3\linewidth}
  \includegraphics[width=1\linewidth]{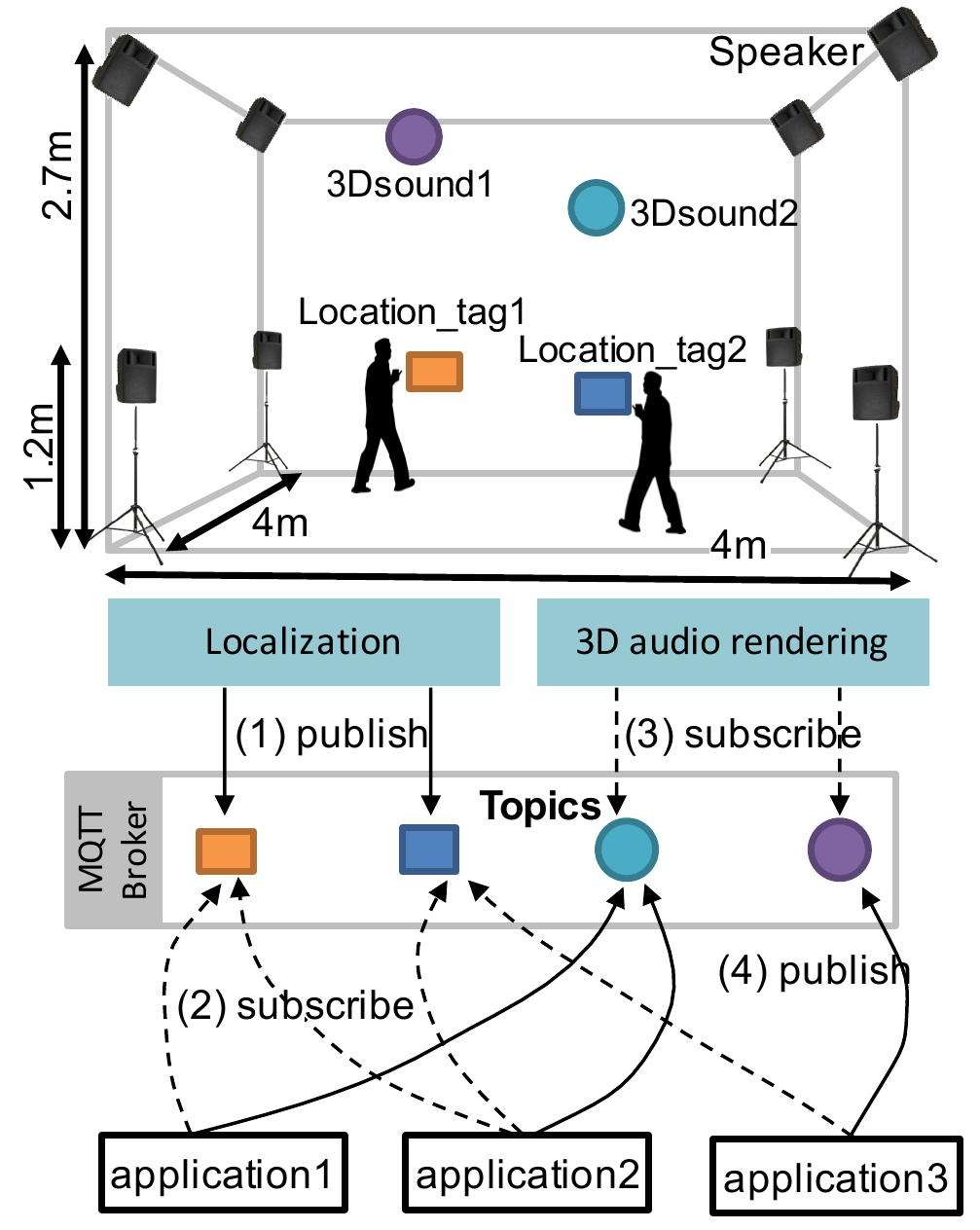}
  \caption{Hardware configuration and access of SDM services}
  \label{fig:mqtt}
  \end{minipage}
\end{figure*}

The \textit{\textbf{application layer}} includes applications that use the SDM services in the SDM station.  One of the applications, called the \textit{reproduction manager}, manages image and sound reproduction based on audience demand with real-time and achieved content from the \textit{public mixer} on the Internet.
The public mixer preprocesses the real-time and archived content, e.g., by collaborative editing.
The \textit{media transmitter} defined in the application layer enables an SDM station to operate as a broadcast station for broadcasting to remote locations. The broadcast from one SDM station is transmitted to other SDM stations as well as to the \textit{content management system} on the Internet.

\section{SDM Prototype System}\label{sec:prototype}

We constructed an SDM station prototype system based on the SDM architecture in the I-REF Building at the University of Tokyo.
%
% \subsection{Overview and Design of Prototype}
Fig.~\ref{fig:prototype-system} shows an overview of the prototype system.
%\begin{figure*}[ht]
%
%  \centering
%
%  \includegraphics[scale=0.5]{figures/prototype-system.pdf}
%
%  \caption{Prototype systems of SDM}
%
%  \label{fig:prototype-system}
%
%\end{figure*}
%

An IP network interconnects the SDM server, the users’ PCs (clients) with the SDM applications, and the devices (localization sensors, speakers, and microphones).
The SDM server manages the 3D locations of the audio-visual objects in the virtual space.
The SDM server runs Windows and provides SDM services (localization, 3D audio rendering, audio extraction) to the applications via the SDM API. All the SDM services are installed in a single server in this prototype, although
they can be distributed to multiple servers.

First,
%we describe the network configuration and the communication protocols in Section~\ref{subsec:network}. Then,
we explain the network configuration and application development in Section~\ref{subsec:sdm-api}. Then, we discuss the SDM services, namely indoor localization, 3D audio rendering, and audio extraction, in Section~\ref{subsec:localization}, Section~\ref{subsec:3d-audio}, and Section~\ref{subsec:audio-extraction}, respectively. The integration of the audio extraction service
 %and video extraction service
 with the SDM prototype system will be investigated in future work.

\subsection{Network Configuration and Development Environment}\label{subsec:sdm-api}
The SDM applications access the server and other SDM applications using Message Queuing Telemetry Transport (MQTT)~\cite{ISO-IEC-20922}, which is developed as a publish/subscribe messaging transport protocol for communication in IoT.
We configure an MQTT broker in the Local Area Network (LAN) and another MQTT broker in the cloud. The MQTT brokers can be switched with each other by the MQTT configuration.

\begin{figure}[ht]
  \centering
  \includegraphics[width=1\columnwidth]{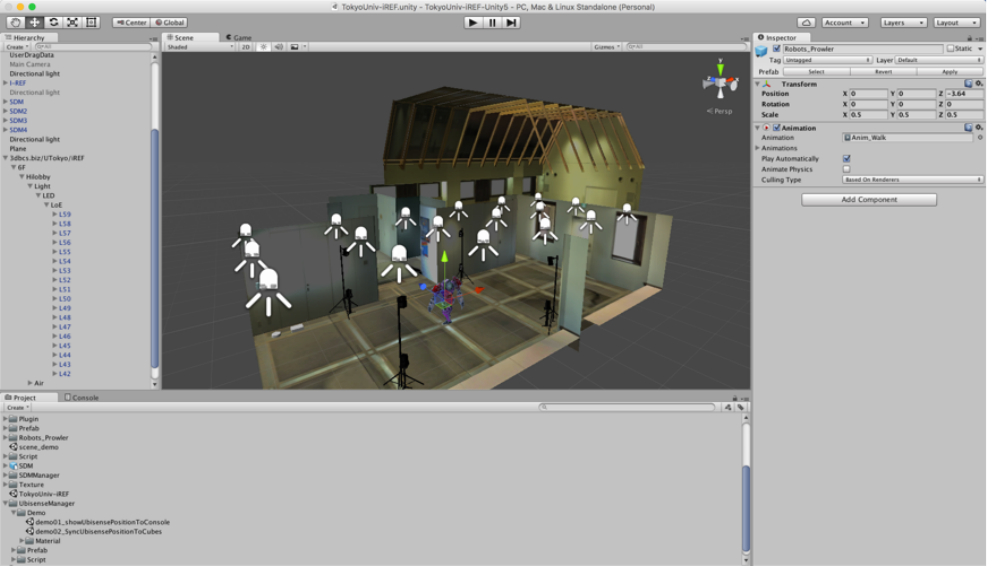}
  \caption{Development environment}
  \label{fig:development-environment}
\end{figure}

SDM applications are developed using Unity\footnote{\url{https://unity3d.com/}}, which can run on Windows, MAC OSX, and Linux. We provide a Unity asset to access the SDM API using MQTT as a sample code. Further, we provide the 3D model of the I-REF building at the University of Tokyo, made using Info360 provided by U's Factory\footnote{\url{http://us-factory.jp/robot/}}. Fig.~\ref{fig:development-environment} shows the Unity application development environment with the import of the Unity assets and the 3D model.

At present, the localization services using Ubisense (detailed in Section~\ref{subsec:localization}) and the 3D audio rendering using VSSS and TH-S (detailed in Section~\ref{subsec:3d-audio}) are integrated into the development environment.
As shown in the Fig.~\ref{fig:mqtt}, the objects managed in the virtual space are associated with topics of MQTT (\textit{e.g., /UTokyo/IREF/Location\_tag1}).
The application can obtain each tag location and the status of each 3D sound at any time, owing to the allocation of individual MQTT topics to each tag and each 3D sound.

%/Users/tsukada/Desktop/TangibleSound/SmartlifeHackathon-iREF1212_updateByTakkin/Assets/SDMManager
%  6                 public static readonly string TOPIC_BASE_VSSS = "3dbcs.biz/UTokyo/iREF/6F/Hilobby/SDM/midi/sc/VSSS01/";
%  7                 public static readonly string TOPIC_BASE_THS = "3dbcs.biz/UTokyo/iREF/6F/Hilobby/SDM/osc/sc/THS/";
% 3dbcs.biz/UTokyo/iREF/6F/Hiloby/Ubisense/id/

\subsection{Indoor Localization Service}\label{subsec:localization}
The indoor localization service uses Real-Time Location System (RTLS) of Ubisense\footnote{\url{http://ubisense.net/}} via ultra-wideband (UWB).
The system uses a small wireless tag (weight, 25 g; dimensions, $1.5\times 1.5\times 0.65$ cm) that emits wireless signals having a frequency of $8.5 \sim 9.5 GHz$ (bandwidth, 1 GHz). The signal is received at four sensors that are mounted at high positions in a room and used to estimate the real-time locations of the tags.
The locations are calculated within an error of $15\sim30$ cm by the angle of arrival and time of arrival of the wireless signal in at least two sensors that are precisely synchronized.
As shown in Fig.~\ref{fig:prototype-system}, communication via UDP and TCP is established between the localization sensors and the localization software (Location Engine).

As shown in (1) of Fig.~\ref{fig:mqtt}, the locations of the tags are published to each MQTT topic associated with each tag.
The application can obtain the locations of the desired tags by subscribing to the associated topics ((2) of Fig.~\ref{fig:mqtt}).
Further, Ubisense software can detect push events of a button on each tag. The push event is also published to the associated topic and delivered to the subscribers.

\subsection{3D Audio Rendering Service}\label{subsec:3d-audio}
The 3D audio rendering service uses Virtual SoundScape System (VSSS) developed by Bandai Namco Studios Inc. and Theater Surround (TH-S)developed by Yamaha Corporation.
An application can use either of them by sending an MQTT message depending on the target sound.
During the startup of the 3D audio rendering service, the service preconfigures the registered 3D sound sources and subscribes to the topics associated with each sound ((3) of Fig.~\ref{fig:mqtt}).
The application can publish the volume, 3D location, pitch, play command, and stop command to the associated topic for controlling the target 3D sound ((4) of Fig.~\ref{fig:mqtt}). In addition, the application can obtain the status of the 3D sound of interest by subscribing to the topics.

The processing results of VSSS and TH-S (each channel output to the speaker) are transmitted to Yamaha XMV8280-D via
Dante\texttrademark, which is a technology for digital media networking over IP developed by Audinate Pty. Ltd.
The SDM server uses Dante Virtual SoundCard, which is a software version of the Dante interface, for the transmission and reception of Dante packets.
Then, Yamaha XMV8280-D converts the received IP packets into amplified analog outputs for the speakers.
At present, eight speakers are mounted in the I-REF building. They are fixed in an area of $4\times 4$ m by speaker stands at heights of 1.2 m and 2.7 m (Fig.~\ref{fig:mqtt}).

\subsubsection{Virtual SoundScape System (VSSS)}
VSSS is a state-of-the-art audio rendering system developed by Bandai Namco Studios Inc. for events or studios.
This software applies sound scene libraries used in commercial game products of Bandai Namco group to a multi-channel system so that the audience can enjoy sound scenes in the real world just as they would be experienced in the game world.
The system also provides a highly interactive interface that was originally developed to react to the game player's feedback.

\subsubsection{Theater Surround (TH-S)}
TH-S is a 3D surround panner system developed by Yamaha Corporation for sound and image control in theatrical applications such as dramas and musicals.
The system allows a sound designer to control the positioning of 3D sound in a theatre. This software runs on Windows and uses a panning algorithm for the input signal, so that the user can locate a sound source arbitrarily in the 3D sound space.

\subsection{Audio Extraction Service}\label{subsec:audio-extraction}

%The Intelligent microphone technique~\cite{Niwa2013, Niwa2014a} developed by NTT Media Intelligence Laboratory realise the audio extraction service.
%First, the analog inputs from the microphone array are converted in Yamaha Ri8-D to IP packets of Dante format and sent to the SDM server (Fig.~\ref{fig:prototype-system}).
%Then, the intelligent microphone technique records the target sounds in distance clearly in noisy environment.
%For clear separation of the target sound, the technique first reveals the feature of the target by the microphone array, and then, extracts the sound that matches to the feature.
%The signal processing such as Beamforming and Wiener filter is applied to the observed signal from the microphone array.
%Finally, the target sound are sent to the users headphone by the conversion from the Dante packet to the analog signal via Yamaha XMV8280-D.

The ``intelligent microphone technique''~\cite{Niwa2013, Niwa2014a} developed
by NTT Media Intelligence Laboratory realizes the audio extraction
service. First, using Yamaha Ri8-D, the analog inputs from the microphone
array are converted into IP packets of Dante
format and sent to the SDM server (Fig.~\ref{fig:prototype-system}). By applying the intelligent
microphone technique, it is possible to emphasize the target sound by suppressing the surrounding noise. For clear enhancement of the target sound, a technique for identifying
the target is performed; then, the
target sound is extracted by suppressing the output levels of the surrounding noise. The signal processing step involves
beamforming and Wiener post-filtering. Finally, the enhanced sound signals are
transmitted to the user’s headphones via conversion of the Dante
packets into analog signals using Yamaha XMV8280-D.

\section{SDM Applications and Demonstration}\label{sec:applications}

\begin{figure}[ht]
  \centering
  \includegraphics[width=1\columnwidth]{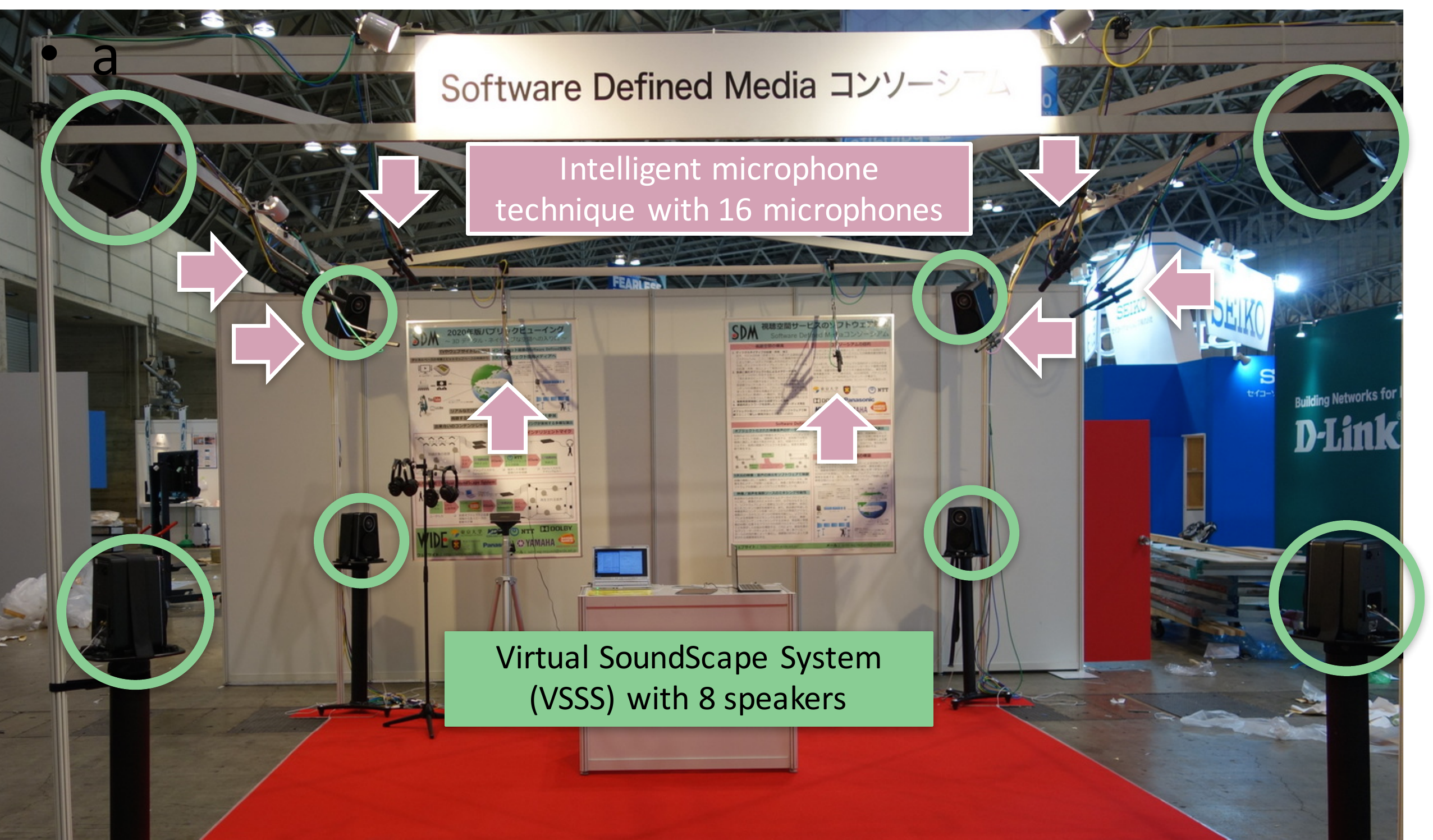}
\caption{Demonstration at Interop Tokyo 2015}
\label{fig:interop2015}
\end{figure}

The SDM consortium presented our prototype system at an exhibition, i.e., Interop Tokyo 2015, held for three days from June 10, 2015, in Makuhari Messe, Chiba, Japan.
The exhibition was attended by 136,341 visitors, most of whom were professionals in the fields of information systems, network engineering, sales, and research.
The presented system was a subset of the SDM prototype described in the previous section. Fig.~\ref{fig:interop2015} shows the system demonstrated at the event. We presented the 3D audio rendering service using VSSS (Section~\ref{subsec:3d-audio}) and the audio extraction service using the intelligent microphone technique (Section\ref{subsec:audio-extraction}).

Some modification was made for the demonstration so that the users could interact with the two services via the GUI at the SDM server instead of accessing the SDM API via MQTT.
For the 3D audio rendering service, eight speakers were installed in
an area of $4\times 4 m$ using speaker stands at heights of 1.2 m and 2.7 m.
The users could interactively control eight sound scenes (including flying mosquito, running dog, robot footstep, and fireworks) by touching an Apple iPad connected to the SDM server via MIDI.
For the audio extraction service, 16 microphones were mounted at eight points at a height of 2.5 m. Each point had two directional microphones, one facing the inside of the booth and the other facing the outside to cancel the outside noise.
The users could clearly hear conversations in one of the four areas that they specified in the GUI of the SDM server.

\begin{figure}[ht]
  \centering
  \includegraphics[width=1\columnwidth]{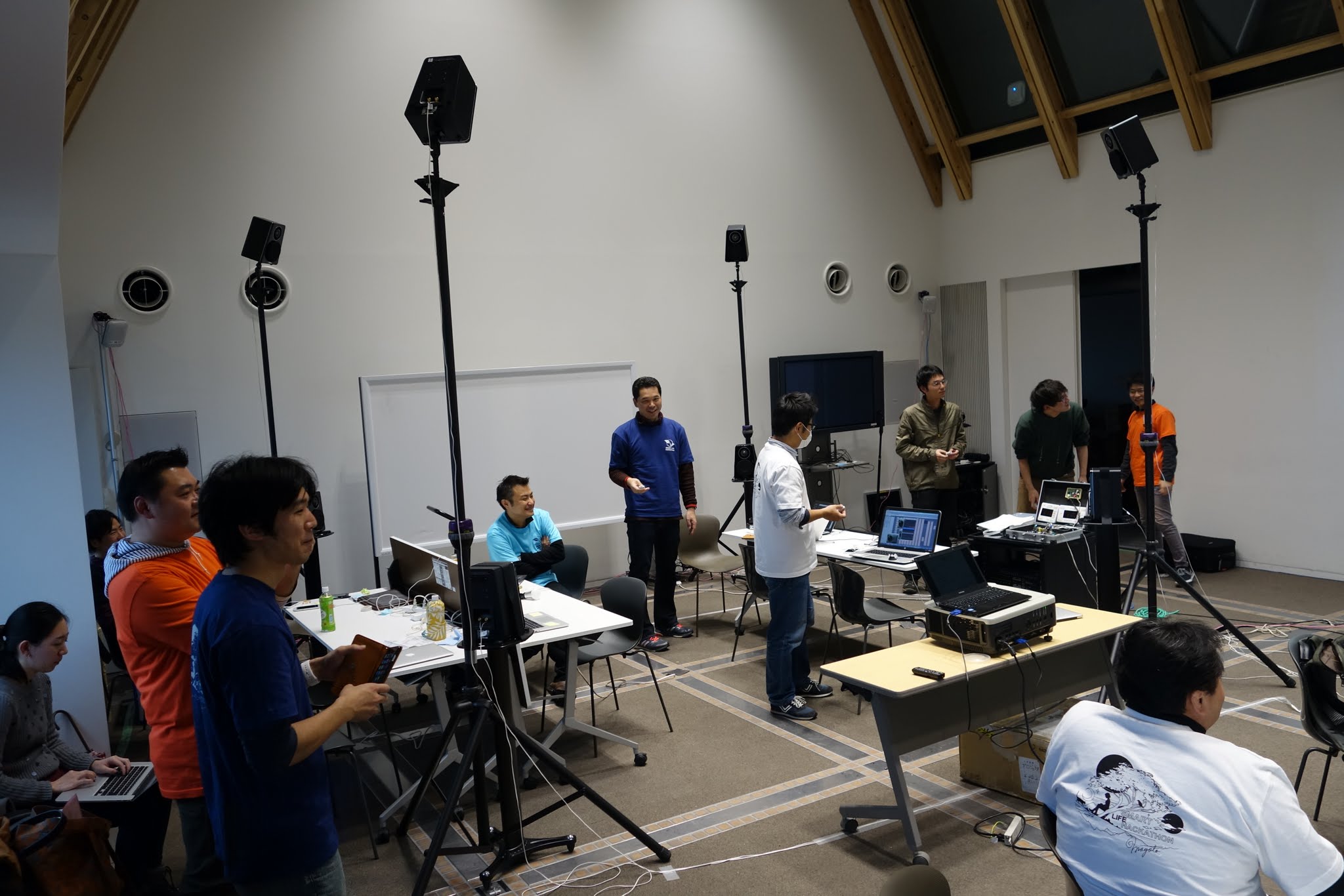}
\caption{Hackathon with SDM platform}
\label{fig:hackathon}
\end{figure}

We hosted two hackathons on the weekends of July 18--19 and December 12--13, 2015, in the I-REF building at the University of Tokyo. The SDM prototype was provided along with other smart building technologies based on IEEE 1888~\cite{IEEE1888-2016}, which were developed for energy management as well as lighting and HVAC control in buildings.
The application development environment and Unity asset sample code for accessing the SDM API described in Section~\ref{subsec:sdm-api} were openly available to the application developers.  The objectives of the hackathons were to develop innovative SDM applications, investigate how the SDM API is used, and collect feedback on the SDM prototype.
Fig.~\ref{fig:hackathon} shows the SDM prototype system provided at the second hackathon.

In the first hackathon, 28 attendees formed 13 teams, and 13 applications were presented.
We only provided the 3D audio rendering service using VSSS for the SDM prototype as well as some sample codes, indicating that the SDM platform had adequate extensibility and flexibility.
Responses to questionnaires circulated after the hackathons indicated that the attendees were "very satisfied" or "satisfied" with the results (responses collected from 10 attendees). Moreover, according to 60\% of the responses, the SDM prototype was the most interesting technology among the provided smart building APIs.
In the second hackathon, 19 attendees developed 9 applications. We provided the 3D audio rendering using TH-S and the indoor localization service using Ubisense in addition to the 3D audio rendering using VSSS. The newly introduced indoor localization service was extremely popular; it was used in all the applications.

%\begin{figure}[ht]
%\begin{minipage}{0.49\columnwidth}
%\centering
%\includegraphics[width=1.7in]{figures/interop2015.pdf}
%\caption{Demonstration of Interop 2015}
%\label{fig:interop2015}
%\end{minipage}
%\begin{minipage}{0.49\columnwidth}
%\centering
%\includegraphics[width=1.7in]{figures/hackathon.pdf}
%\caption{Hackathon with SDM platform}
%\label{fig:hackathon}
%\end{minipage}
%\end{figure}

\section{Evaluation} \label{sec:evaluation}

This section describes the experimental evaluation of the SDM prototype described in Section~\ref{sec:prototype}.

\subsection{Sound Reproducibility Test}

A sound reproducibility test was conducted to study the reproducibility of 3D sound produced by the prototype system.
% the author explored how users feel and experience the 3-D sound which played from the system of Tangible Sound Object as pre-experiment.
In the experiment, we investigated how users estimate the locations of the sound objects and the distance between themselves and the sound objects by listening to the sounds produced by the system.

Responses were collected from 12 participants, comprising an equal number of males and females ranging in age from the teens to the fifties.
The experiment was conducted at the University of Tokyo, and the settings of all the systems were identical to those of the second hackathon. Each participant was positioned at the center of the designated space surrounded by speakers and given a questionnaire to provide responses/comments about his/her experience after listening to 3D sounds produced by the prototype system.

The sounds used in the experiment are summarized in Table~\ref{table:typesofsounds}. Sounds V1-V4 were processed by VSSS and sounds T5-T6 were controlled by Th-S as the 3D audio rendering service.  All the sounds except for V2 and V3 had 3D expression including height expression. On the other hand, the dog and robot sounds (V2 and V3) did not have height expression, because these sounds were designed to move along the ground.
All the sounds were played more than twice; first, the sounds were static, and then, the sounds were dynamic. The sounds were played in random order and the locations of sounds were random as well. The sounds, locations, and movements were controlled via Unity.

The questions were as follows:
\begin{itemize}
  \item Q1) What was the sound you heard?
  \item Q2) Where was the location of the sound object? (right, left, front, back...)
  \item Q3) Could you follow the track of sound movement? (rate on a scale of 1 to 10)
\end{itemize}

The answers to question Q1 were very clear, all the participants could recognize all the sounds. The answers to Q2 and Q3 are listed in Table~\ref{table:typesofsounds}.

%\begin{table}[ht]
%\caption{Types of Sounds and Result of Experiment}
%\label{table:typesofsounds}
%\begin{center}
%\small
%\scalebox{0.95}{
%\begin{tabular}{|c|c|c||c|c|}
%\hline
%No. &Sound &3D Expression &Q2) & Q3) \\
% & & & Location & Movement \\
%\hline
%\hline
%V1  & Mosquito    & YES  & 70\% & 6 \\
%\hline
%V2 & Dog  & On the ground & 70\% & 3 \\
%\hline
%V3 & Robot footstep & On the ground & 80\% & 8 \\
%\hline
%V4 & Firework  & YES  & 80\% & 8 \\
%\hline
%\hline
%T5  & Song voice & YES & 90\% & 7 \\
%\hline
%T6  & Piano   & YES  & 80\% & 6 \\
%\hline
%\end{tabular}
%}
%\end{center}
%\end{table}

\begin{table}[ht]
\caption{Types of sounds and results of the experiment}
\label{table:typesofsounds}
\begin{center}
\small
\scalebox{1}{
\begin{tabular}{ccccc}
\toprule
No. &Sound &3D Expression &Q2) & Q3) \\
 & & & Location & Movement \\
\hline
V1  & Mosquito    & YES  & 70\% & 6 \\
V2 & Dog  & On the ground & 70\% & 3 \\
V3 & Robot & On the ground & 80\% & 8 \\
V4 & Firework  & YES  & 80\% & 8 \\
\hline
T5  & Song voice & YES & 90\% & 7 \\
T6  & Piano   & YES  & 80\% & 6 \\
\bottomrule
\end{tabular}
}
\end{center}
\end{table}

When the sounds were static (Q2), the locations that most of participants could recognize were front, behind, right, left, diagonally in front, and diagonally behind. The participants could estimate the locations of the sounds correctly in 70\%--90\% of the cases.
To recognize the location at a height, it is necessary to compare sounds from above and below. For example, to hear a sound from above, it is necessary to listen to the sound below before listening to it from above. It was possible to roughly follow the dynamic sounds. The firework sound (V4) was the most suitable sound for expressing the height. This was attributed to the height of the original sounds.

When the sounds were dynamic (Q3), the sounds that were not continuous were difficult to follow, especially the dog sound (V2).
The duration of the firework sound was not so long, but the sound was moving from the ground to the air (long distance) continuously within a few seconds, which could be heard very clearly by the participants. The movement of the song voice sound (T5) could also be followed well, except for the gaps between the phrases of the songs.
According to the comments of the participants, other sounds having height expression, such as song voice (T5) and piano (T6), seem interesting when they move up and down because such sound effects are not experienced in real life.

\subsection{Real-timeness Test}
We performed access delay measurement of the SDM API with MQTT because the SDM services need interactiveness for the applications.
In the prototype network configuration where the application and the server are connected to the same LAN, the delay of a message published from an application to the SDM server via the broker is equal to the round-trip time (RTT) between the application and the broker, when the application and the server are subscribing to the same topic.
Therefore, we can measure the messaging delay of SDM API access using MQTT by measuring the messaging delay between the application and the broker.
We measure the RTT with various settings, including message frequency, message size, and location of the broker.
Table~\ref{tab:System-configuration} summarizes the system configuration of the local broker, cloud broker, and client PC.

\begin{table}[ht]
\caption{System configuration\label{tab:System-configuration}}
\centering{}
\begin{tabular}{p{0.8cm}p{1.3cm}p{1.65cm}rp{2.2cm}}
\toprule
 & OS & CPU & Memory & Software\tabularnewline
\midrule
Local Broker & Ubuntu 14.04 LTS & Intel\textregistered Core\texttrademark   \  \      2 Duo E6600 CPU 2.40 GHz & 5.8 GB & Mosquitto (ver 1.4.2 MQTT v3.1 broker)\tabularnewline
\hline
Cloud Broker & Ubuntu 14.04 64bit small & 1vCPU & 1 GB & Mosquitto (ver 1.4.4 MQTT v3.1 broker)\tabularnewline
\hline
PC & Windows 7 Enterprise & Intel\textregistered Core\texttrademark \  \    i7-3630QM CPU 2.40 GHz & 16.0 GB & Unity 5.3.0\tabularnewline
\bottomrule
\end{tabular}
\end{table}

\begin{figure}[ht]
  \centering
  \includegraphics[width=1\columnwidth]{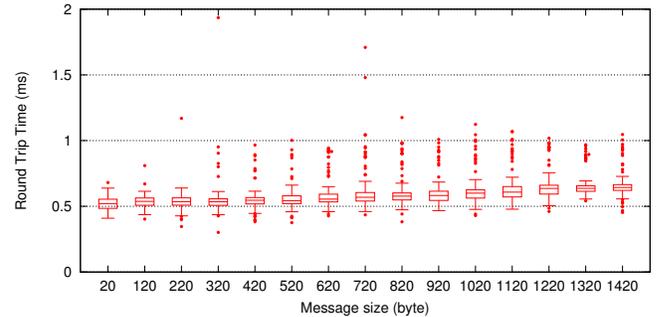}
  \caption{RTT of MQTT messages with varying message size}
  \label{fig:mqtt-rtt-local-size}
\end{figure}

\begin{figure}[ht]
  \centering
  \includegraphics[width=1\columnwidth]{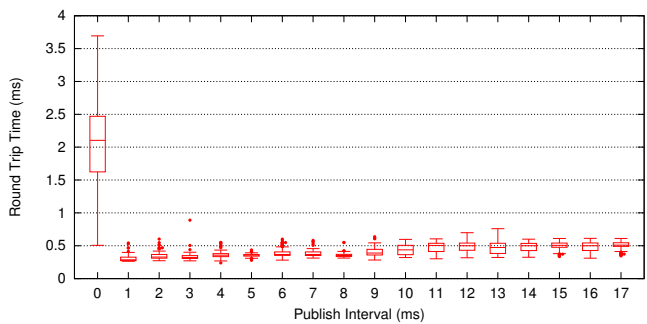}
  \caption{RTT of MQTT messages with varying message frequency}
  \label{fig:mqtt-rtt-local-freq}
\end{figure}

\begin{figure}[ht]
  \centering
  \includegraphics[width=1\columnwidth]{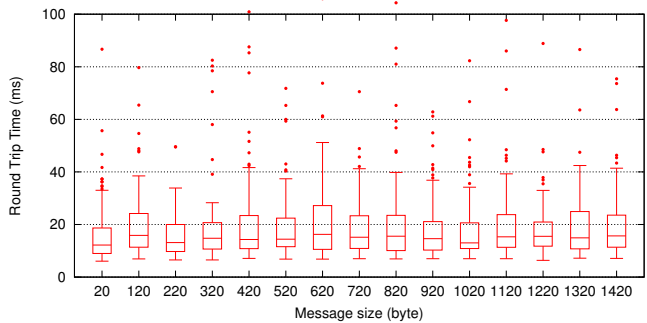}
  \caption{RTT of MQTT messages with cloud broker}
  \label{fig:mqtt-rtt-cloud}
\end{figure}

Fig.~\ref{fig:mqtt-rtt-local-size} shows the RTT between the application and the broker when the broker was connected to the same LAN. The delay of each message size was measured 100 times, where the interval of the message was 17 ms. The interval was equivalent to the access to the SDM API by the Unity application once in a display refresh (the normal frame rate of unity is 60 frames per second).  To avoid the influence of fluctuation of processing resources in the client and broker in a certain period, the MQTT packet was transmitted with all packet sizes sequentially and this was repeated 100 times.
The messages were delivered in the range of 0.4--0.75 ms when the message size was changed from 20 bytes to 1420 bytes. As the message size increases, the RTT becomes slightly longer. The median of the RTT was 0.52 ms and 0.65 ms for packets of 20 bytes and 1420 bytes, respectively.
The boxplots in Fig.~\ref{fig:mqtt-rtt-local-size}-\ref{fig:mqtt-rtt-cloud} show the maximum, third quartile, first quartile, and minimum RTT.

Fig.~\ref{fig:mqtt-rtt-local-freq} shows the RTT of MQTT messages of 60 bytes with various frequencies from 0 ms to 17 ms. From the results, we can confirm that the message delay of SDM API access was around 0.5 ms when there were 17 applications in the same LAN. Further, the delay of 0 ms in Fig.~\ref{fig:mqtt-rtt-local-freq} represents the case in which the messages were sent without an interval. In this case, the median of the RTT was around 2 ms.

%\begin{figure}[ht]
%  \begin{minipage}{0.49\columnwidth}
%  \centering
%  \includegraphics[width=1.7in]{figures/mqtt-rtt-local-size.pdf}
%  \caption{RTT of MQTT messages when message size changes}
%  \label{fig:mqtt-rtt-local-size}
%  \end{minipage}
%  \begin{minipage}{0.49\columnwidth}
%  \centering
%  \includegraphics[width=1.7in]{figures/mqtt-rtt-local-freq.pdf}
%  \caption{RTT of MQTT messages when message frequency changes}
%  \label{fig:mqtt-rtt-local-freq}
%  \end{minipage}
%\end{figure}

On the other hand, Fig.~\ref{fig:mqtt-rtt-cloud} shows boxplots of RTTs with various message sizes in the configuration where the MQTT broker was in the cloud.
The figure shows that the most of packets returned within 50 ms; moreover, around 80\% of the messages returned within 30 ms.
This indicates that the delay of SDM API access in the cloud broker configuration was within 2 or 3 frame refreshes.
We could not identify the difference in RTT by the message size. The sub-millisecond-order difference by the message size found in the results with the broker in the same LAN (Fig.~\ref{fig:mqtt-rtt-local-size}) did not appear in this case. In comparison with the local broker, we found more extreme outliers in the results with the cloud broker. Specifically, 37 MQTT messages did not return within 100 ms and 2 MQTT messages did not return within 500 ms out of a total of 1500 messages. We need to investigate the reason for the outliers with the cloud service operators in the future.

\section{Conclusion and Future Work}\label{sec:conclusion}
This paper discussed the activities of the SDM consortium that we had established in 2014 to target new research areas and markets involving object-based digital media and Internet-by-design audio-visual environments.
We proposed the SDM architecture that virtualizes the networked audio-visual infrastructure and provides flexible access of audio-visual services to SDM applications on the basis of SDN. Then, we constructed the proof-of-concept implementation based on the architecture. A prototype SDM station system was presented at an exhibition held in Japan and provided to SDM application developers at two hackathons. User experience tests showed that the prototype provides effective 3D audio reproducibility. Furthermore, the evaluation of SDM API access showed that it maintains the interactiveness of the SDM applications in most of the cases.

We are considering three directions for future work.
First, we plan to implement the SDM API of the intelligent microphone technique using MQTT, by replacing the GUI access to the SDM server.
Second, we plan to use a free-viewpoint video generation system~\cite{Sankoh2012} developed by an SDM consortium member,
namely KDDI R\&D Laboratories, Inc., for the video extraction service.
Third, we plan to not only continue updating the SDM station prototype system but also construct other prototypes (e.g., public mixer, content management).
For content management, we plan to study a flexible audio-visual database for application development using recorded concert data detailed in~\cite{Ikeda2016}.


\section*{Acknowledgment}
The authors would like to thank the SDM consortium members.

%\section*{TODO}
%- [words] \\
%-- audio extraction\\
%-- 3D audio reproduction $\rightarrow$ 3D audio rendering\\
%-- Local renderer $\rightarrow$ rendering manager\\
%-- Public renderer $\rightarrow$ public mixer\\
%\\
%- [more detail]\\
%--- ``by taking the hint from Software Defined Networking (SDN) advanced rapidly in the networking development.''\\
%
%--- hardware configuration in prototype\\
%------ how many channel to speaker and microphone?
%
%- [contents]\\
%---- future works: free-viewpoint video generation system~\cite{Sankoh2012} developed by KDDI R\&D Laboratories

\bibliographystyle{IEEEtran}
\bibliography{sdm-report.bib}

\end{document}